\documentclass[12pt]{article}
\usepackage{a4wide}
\usepackage{amssymb}
\usepackage{graphicx}
\begin{document}
{\renewcommand{\thefootnote}{\fnsymbol{footnote}}
\medskip
\begin{center}
{\LARGE  Quantum gravity effects on space-time}\\
\vspace{1.5em}
Martin Bojowald\footnote{e-mail address: {\tt bojowald@gravity.psu.edu}}
\\
\vspace{0.5em}
Institute for Gravitation and the Cosmos,\\
The Pennsylvania State
University,\\
104 Davey Lab, University Park, PA 16802, USA\\
\vspace{1.5em}
\end{center}
}

\setcounter{footnote}{0}

\begin{abstract}
 General relativity promotes space-time to a physical,
  dynamical object subject to equations of motion. Quantum gravity,
  accordingly, must provide a quantum framework for space-time,
  applicable on the smallest distance scales. Just like generic states
  in quantum mechanics, quantum space-time structures may be highly
  counter-intuitive. But if low-energy effects can be extracted, they
  shed considerable light on the implications to be expected for a
  dynamical quantum space-time. Loop quantum gravity has provided
  several such effects, but even in the symmetry-reduced setting of
  loop quantum cosmology no complete picture of effective space-time
  geometries describing especially the regime near the big bang has
  been obtained. The overall situation regarding space-time structures
  and cosmology is reviewed here, with an emphasis on the role of
  dynamical states, effective equations, and general covariance.
\end{abstract}

\section{Introduction}

In modern cosmology, according to the common scenarios, one is using
the universe and its own expansion as a microscope aimed at the smallest
distance scales of space and time. In order to understand the
resulting phenomena to be expected, the nature of the correct
microscopic degrees of freedom, out of which space-time and its
dynamics is to emerge, must be understood. There are no direct
observations to guide us, and so we are required to make use of
further input, of principles that are strong enough to support a large
theoretical edifice.

A well-known example for the theoretical derivation of new microscopic
degrees of freedom from an underlying principle is the electroweak
theory. One of its motivations was a well-defined description of
$\beta$-decay, a reaction of four particles (seen at least via the
energy carried away in the case of the neutrino). As a pointlike
interaction between the four particles involved, the perturbative
quantum field theoretical description does not lead to well-defined
decay rates. A new principle, renormalizability, is used to look for a
more suitable theoretical framework. Its implementation requires the
inclusion of new quantum degrees of freedom, the exchange bosons of
the electroweak theory. Based on the principle of renormalizability,
they were predicted theoretically well before direct observations by
particle accelerators became possible.

In quantum gravity, we are currently in a situation similar to that
before the direct detection of the exchange bosons. We do not have
direct evidence for the microscopic degrees of freedom of quantum
gravity, but we do know several problems of the classical theory,
chiefly the singularity problem of general relativity. Again, extra
input for a theoretical description is needed in the form of
principles. One may decide to use those already tried and true, such
as renormalizability in this case leading to string theory (as per
current understanding). In this way, a quantization of gravitational
(and other, unified) excitations on space-time becomes possible. There
are, however, difficulties in the description of strong gravitational
fields, as we find them at the big bang or in black holes. The
interaction of matter with the space-time structure is relevant in
those regimes, and so we must consider the quantum nature of full
space-time. While this is not impossible in string theory, the
theory's setup makes the analysis of such questions rather indirect.
 
An alternative principle offers itself, based on what we know about
regimes of strong gravitational fields: the principle of background
independence. It states the requirement of quantizing the full metric
$g_{\mu\nu}$ known as the representative of space-time geometry as it
must emerge at large distances or low energies; it is not enough to
just quantize perturbations $h_{\mu\nu}$ on a given (e.g.\
Minkowskian) background where $g_{\mu\nu}=\eta_{\mu\nu}+h_{\mu\nu}$.
If only $h_{\mu\nu}$ is quantized and $\eta_{\mu\nu}$ kept as a
classical metric background, the theory does not describe the complete
space-time geometry in a quantized way. There will be physical quantum
degrees of freedom for $h_{\mu\nu}$, but a classical, rigid background
$\eta_{\mu\nu}$ remains in the theory. It may be possible to find
observables insensitive to which $\eta_{\mu\nu}$ is used, but this
would be difficult to achieve and to demonstrate. Moreover, regimes of
strong gravitational fields, especially near classical singularities,
do not allow a perturvative treatment with a small $h_{\mu\nu}$: The
metric becomes degenerate, and so at least some of the crucial
components of $h_{\mu\nu}$ are as large as those of $\eta_{\mu\nu}$.
Here, a background independent quantization of the full $g_{\mu\nu}$
becomes most useful, a treatment realized in loop quantum gravity.

\section{Background independence}

Quantum field theory on a background Minkowski space-time may be
formulated by operators $a_{\bf k}$ and $a_{\bf k}^{\dagger}$ that
describe the annihilation and creation of particles of momentum ${\bf
  k}$.  Using $a_{\bf k}^{\dagger}$ introduces a new particle and
increases the total energy, while products of operators in a
Hamiltonian amount to interactions.  One problem to be faced in
quantum gravity is that particles can only be created on a given
space-time, whose metric is used in the very definition of $a_{\bf k}$
and $a_{\bf k}^{\dagger}$.  A possible solution is to define operators
for space-time itself.  Such operators would increase distances, areas
and volumes; not energy.

Loop quantum gravity \cite{Rov,ThomasRev,ALRev} provides a
specific realization of this solution, at least for spatial geometry.
Creation operators turn out to be holonomies \cite{LoopRep}
$h_I={\cal P}\exp(\smallint_{e_I}{\rm d} t \dot{e}_I^aA_a^j\tau_j)$
along spatial curves $e_I$, evaluated for a connection
$A_a^j=\Gamma_a^j+\gamma K_a^j$ called the Ashtekar--Barbero
connection \cite{AshVar,AshVarReell}. The structure group of the
theory is SO(3), referred to by the index $j$ and geometrically
corresponding to local spatial rotations. (The matrices
$\tau_j=-\frac{1}{2}i\sigma_j$ are generators of su(2), proportional
to the Pauli matrices $\sigma_j$.)  In the definition of the
connection, we use the spatial spin connection $\Gamma_a^i$ and
extrinsic curvature $K_a^i$, with the Barbero--Immirzi parameter
$\gamma>0$ \cite{AshVarReell,Immirzi}, whose value parameterizes
a family of canonical transformations.  Classically, the connection is
canonically conjugate to a densitized vector field $E^a_i$, the
densitized triad related to the spatial metric $q_{ab}$ by
$E^a_iE^b_j=\det q q^{ab}$: $\{A_a^i(x),E^b_j(y)\}= 8\pi\gamma G
\delta_a^b\delta^i_j\delta(x,y)$. This field determines the
(torsion-free) spin connection
\begin{equation}\label{SpinConn}
 \Gamma_a^i= -\epsilon^{ijk}e^b_j
(\partial_{[a}e_{b]}^k+ {\textstyle\frac{1}{2}}
e_k^ce_a^l\partial_{[c}e_{b]}^l)\,.
\end{equation}
Only the extrinsic curvature part
of $A_a^i$ is independent of the triad.

To start setting up the connection representation for a quantum
formulation, we define a basic state $|0\rangle$ by $\langle
A_a^i|0\rangle=1$, fully independent of the connection.  A basis of
states is obtained by using holonomies as creation operators, which in
the simplified U(1)-example (where $h_e= \exp(i\int_e{\rm d}t
\dot{e}^aA_a)$) can be written as
\[
 |e_1,k_1;\ldots;e_I,k_I\rangle=
\hat{h}_{e_1}^{k_1}\cdots \hat{h}_{e_I}^{k_I}|0\rangle\,.
\]
Similar, though more tedious formulas apply for the SU(2)-case of
quantum gravity, whose states can be expanded in terms of spin networks
\cite{RS:Spinnet}. A general state is labeled by a graph $g$ with
integers $k_e$ as quantum numbers on its edges $e$:
\[
 \psi_{g,k}(A_a)=\prod_{e\in g}
h_e(A_a)^{k_e}=\prod_{e\in g} \exp(ik_e \smallint_e {\rm d} t \dot{e}^a A_a)\,.
\]

Holonomies $h_I$ create quantum-gravity states by excitations of two
types.  (i) one can use operators several times for the same curve, or
(ii) use different curves. In this way, an irregular lattice, or spin
network, arises which intuitively can be seen as the microscopic
structure of space.  For a macroscopic geometry, a dense mesh, or
strong excitations of the quantum gravity state, are necessary; in
order to model near-continuum geometries for which general relativity
may approximately apply, one has to consider ``many-particle'' states.

In order to extract geometrical notions from the visualization of
states, operators representing the densitized triad must be
introduced. From the densitized triad, the spatial metric and then
usual geometrical quantities result \cite{AreaVol,Area,Vol2}.
As we had to integrate the connection along curves to obtain
holonomies, the densitized triad must be integrated 2-dimensionally in
order to obtain well-defined operators: the fluxes $\int_S{\rm d}^2y
E^an_a$ integrated over spatial surfaces $S$ with the
metric-independent co-normal $n_a$ (again written here for the
U(1)-simplification). Obtained from momenta conjugate to the
connection, fluxes are quantized to derivative operators. Their
specific action shows that they measure the excitation level of a
state along edges intersecting the surface:
\[
  \int_S{\rm d}^2y n_a\hat{E}^a\psi_{g,k}=8\pi \gamma G
\int_S{\rm d}^2y n_a \frac{\hbar}{i}\frac{\delta \psi_{g,k}}{\delta A_a(y)}=
8\pi\gamma\ell_{\rm P}^2\sum_{e\in g} k_e{\rm  Int}(S,e) \psi_{g,k}
\]
with the intersection number ${\rm Int}(S,e)$ and the Planck length
$\ell_{\rm P}= \sqrt{G\hbar}$.  This is an eigenvalue equation with
discrete eigenvalues read off as $8\pi\gamma\ell_{\rm P}^2\sum_{e\in
  g} n_e{\rm Int}(S,e)$, and so spatial geometry, represented by the
fluxes, is discrete in this framework.  The graphs obtained from
elementary excitations represent the atomic nature of space, and
geometry results from intersections.

In order to see how discrete spatial structures of this kind evolve,
dynamics must be introduced. For gravity, this is determined by the
Hamiltonian constraint, the phase space functional
\[
 C_{\rm grav}[N] = \frac{1}{16\pi\gamma G} \int_{\Sigma} \mathrm{d}^3x N
 \left(\epsilon_{ijk}F_{ab}^i\frac{E^a_jE^b_k}{\sqrt{|\det
E|}}  -
2(1+\gamma^{-2})
(A_a^i-\Gamma_a^i)(A_b^j-\Gamma_b^j)\frac{E^{[a}_iE^{b]}_j}{\sqrt{|\det E|}}
 \right)
\]
in Ashtekar variables. In addition to the fields already introduced,
we use the curvature $F_{ab}^i$ of $A_a^i$ and the lapse function $N$.
As a constraint, $C_{\rm grav}[N]$ must vanish for all choices of $N$.

There are several apparent obstacles in turning this expression into
an operator, using the basic holonomies and fluxes. First, an inverse
determinant of the densitized triad is required but seems problematic
at the operator level, where fluxes, with discrete spectra containing
zero, lack densely-defined inverse operators. Nevertheless, as shown
by \cite{QSDI}, the quantity needed can be obtained from a relation
such as
\begin{equation} \label{InvTriad}
 \left\{A_a^i,\int{\sqrt{|\det E|}}\mathrm{d}^3x\right\}= 2\pi\gamma G
 \epsilon^{ijk}\epsilon_{abc} \frac{E^b_jE^c_k}{{\sqrt{|\det E|}}}
\end{equation}
whose left-hand side is free of inverses. The volume operator, made
from fluxes, can be used for $\int\sqrt{|\det E|}\mathrm{d}^3x$,
$A_a^i$ can be expressed in terms of holonomies, and the Poisson
bracket is finally turned into a commutator divided by $i\hbar$.

Similarly, the curvature components $F_{ab}^i$ of the Ashtekar
connection can be expressed in terms of holonomies using identities
such as $s_1^as_2^bF_{ab}^i\tau_i= \Delta^{-1}(h_{\lambda}-1)
+O(\Delta)$, where $\lambda$ is a square loop of small coordinate area
$\Delta$, with tangent vectors $s_I^a$ at one of its corners. Finally,
extrinsic curvature components $K_a^i=\gamma^{-1} (A_a^i-\Gamma_a^i)$,
the most complex expressions in the Hamiltonian constraint owing to
the spin connection (\ref{SpinConn}) as a functional of the
densitized triad, can be obtained from what has already been
constructed:
\[
K_a^i=\gamma^{-1} (A_a^i-\Gamma_a^i)
\propto \left\{\!A_a^i,\!\left\{\int{\rm d}^3x F_{ab}^i 
\frac{\epsilon^{ijk}E^a_jE^b_k}{\sqrt{|\det
E|}},\int{\sqrt{|\det E|}}\mathrm{d}^3x\right\}\!\!\right\}\,.
\]

\begin{figure}[t]
\centering
\includegraphics[keepaspectratio=true,width=10cm]{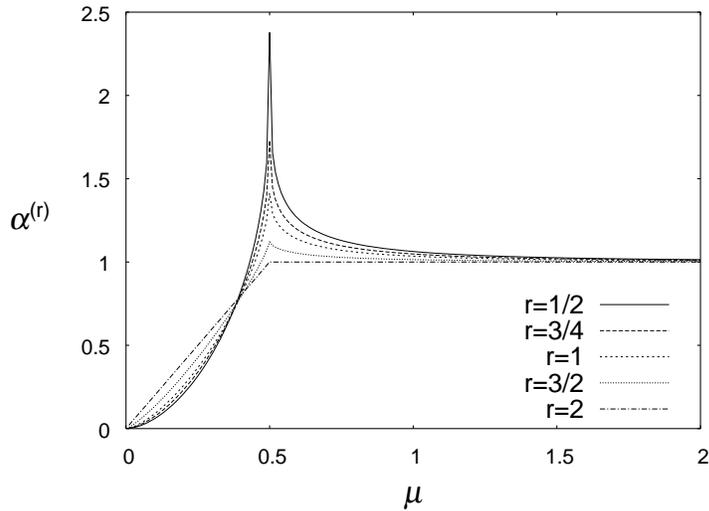}
\caption{Inverse triad correction function, depending on an ambiguity
  parameter $r$. All curves approach the classical expectation one
  from above for large flux values $\mu$. (See
  \cite{Ambig,QuantCorrPert} for explicit derivations of correction
  functions.)  }
\label{fig:alpha05r_to2}
\end{figure}

In this way, a well-defined class of Hamiltonian constraint operators
arises, parameterized by certain ambiguities as they arise in the
choices to be made. Examples for ambiguities are the exact rewriting
of the inverse determinant, or routings and representations for the
holonomies used. In spite of ambiguities, several characteristic
properties can be extracted and evaluated phenomenologically. In
particular, there are three main sources of quantum corrections:
\begin{itemize}
\item Inverse volume corrections arise from quantizing the inverse
  triad determinant in an indirect manner. Comparing eigenvalues of
  the resulting operators quantizing the left hand side of
  (\ref{InvTriad}) with the expressions expected from simply
  inserting flux eigenvalues into the right hand side of
  (\ref{InvTriad}) shows strong deviations for small flux scales;
  see Fig.~\ref{fig:alpha05r_to2}
\item Higher order corrections result from the use of holonomies,
  contributing higher powers of the connection components.
\item As in any interacting quantum field theory, quantum
  back-reaction results from the influence of quantum variables such
  as fluctuations, correlations or higher moments of a state on the
  expectation values. These variables provide extra degrees of
  freedom, which can sometimes be interpreted in the sense of higher
  derivative terms.
\end{itemize}

A simplified form of the Hamiltonian, as it arises from the general
constructions of \cite{RS:Ham,QSDI}, is
 \begin{equation} \label{HamOp}
 \hat{C}_{\rm grav}[N]= 
\frac{i}{16\pi\gamma G\hbar}
\sum_{v,IJK} N(v)
\sum_{\sigma_I\in\{\pm1\}} \sigma_1\sigma_2\sigma_3
\epsilon^{IJK}
{\rm tr}(h_{v,I}
h_{v+I,J}
h_{v+J,I}^{-1} h_{v,J}^{-1}
h_{v,K} [h_{v,K}^{-1},\hat{V}])
\end{equation}
of interacting form: excitations of geometry take place dynamically by
the factors of holonomy operators included in the expression. All this
depends on the spatial geometry through the volume operator $\hat{V}$.
The discreteness contained in the resulting dynamics is significant at
high densities (such as the big bang), or if many small corrections
add up in a large universe (for dark energy, perhaps).

\section{Loop quantum cosmology}

\begin{figure}[t]
\centering
\includegraphics[keepaspectratio=true,width=8cm]{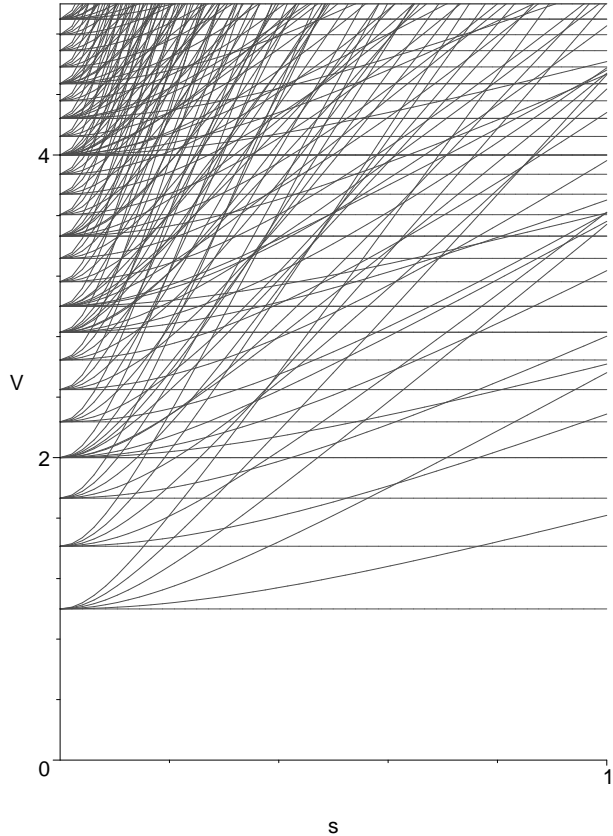}
\caption{Level splitting of volume eigenvalues as a symmetry parameter
  $s$ is tuned from zero (homogeneity) to one (spherical symmetry);
  see \cite{SphSymmVol} for details.  }
\label{fig:Split}
\end{figure}

In full generality, it is difficult to analyze the dynamics of quantum
gravity, but several results are known in model systems (based on
symmetry reduction or perturbative inhomogeneities).  One can easily
imagine simplifications from the considerable reduction of the number
of degrees of freedom, but also from another effect: level-splitting,
well-known from energy spectra of atoms and molecules. Also in quantum
geometry, levels split when symmetries are relaxed, making spectra of
symmetric situations much simpler than non-symmetric ones. In
particular, the volume spectrum, which despite significant numerical
progress \cite{VolSpecI,VolSpecII} is rather difficult to
compute in the full case, splits when symmetry is relaxed from
homogeneity to spherical symmetry as shown in Fig.~\ref{fig:Split}.
The most highly symmetric systems should then the easiest to analyze,
also concerning the dynamics. This is the realm of quantum cosmology.

Loop quantum cosmology \cite{Livrev} provides a quantization of
symmetry reduced models by the techniques of loop quantum gravity.
Many of its ingredients, in particular its states and basic operators,
can be induced from the full holonomy-flux algebra
\cite{SymmRed,SphSymm,InhomLattice,Reduction,Rieffel} or by
other means \cite{SymmQFT,SymmStatesInt}; in this sense loop quantum
cosmology is a sector of (kinematical) loop quantum gravity.  Just the
dynamics, which is problematic and not yet in a settled stage even in
the full theory, is too complicated to be reduced directly from a full
Hamiltonian constraint such as (\ref{HamOp}). In formulating the
dynamics, based on the basic operators, additional assumptions and
extra input are sometimes required and cannot yet be derived from the
full theory. This may introduce an amount of ambiguity larger than
that already realized in the full theory.

Hamiltonian isotropic cosmology in Ashtekar variables (here written
only for spatially flat models), has the basic phase space variables
$A_a^i=c\delta_a^i$ with $c=\gamma\dot{a}$, $E^a_i=p\delta^a_i$ with
$|p|=a^2$. (For a triad with the option of two different orientations,
$p$ can take both signs; see \cite{IsoCosmo} for details of the
classical reduction.) The coefficients $c$ of an isotropic connection
and $p$ of an isotropic triad are canonically conjugate:
$\{c,p\}=8\pi\gamma G/3$. Inserting these reductions into the full
Hamiltonian constaint provides
\[
 C:=-\frac{c^2\sqrt{|p|}}{\gamma^2}+\frac{8\pi G}{3} H_{\rm matter}=0
\]
as the isotropic constraint equation, equivalent to the Friedmann
equation. (For homogeneous cosmology, the lapse function $N$ must be
spatially constant, thus providing a single constraint function.) The
gauge-flow in time, $\dot{p}=\{p,C\}$ and $\dot{c}=\{c,C\}$, generated
by the constraint amounts to the Raychaudhuri equation.

In loop quantum cosmology, just as in Wheeler--DeWitt quantum
cosmology \cite{DeWitt,QCReview}, the constraint is quantized to an
operator annihilating physical states: $\hat{C}|\psi\rangle=0$. In
contrast to the Wheeler--DeWitt representation, however, the use of
$\exp(i\delta c)$ for $c$ --- matrix elements of holonomies as
required for a background-independent representation --- makes us
regularize the constraint before it can be quantized. For instance,
with an ambiguity parameter $\delta$ (which can and often should be
allowed to be a phase-space function; see Sec.~\ref{sec:LattRef})
we may write
\begin{equation} \label{RegHam}
-\frac{\sin^2(\delta
  c)\sqrt{|p|}}{\gamma^2\delta^2}+\frac{8\pi G}{3} H_{\rm
  matter}=0
\end{equation}
as an expression that agrees very well with the classical one for
small curvature ($\delta c\ll 1$) and at the same time is quantizable
in terms of holonomies. Effects of such a modification easily trickle
down to low-curvature equations \cite{EffHam,DiscCorr}.

Replacing connection components by matrix elements of holonomies
constitutes a regularization\footnote{Some models --- including
  cosmological ones with specific matter contents \cite{NonExpLQC},
  parameterized free particle field theories
  \cite{PPFT,PolymerFree} and certain dilaton gravity models
  \cite{PFTLoop} --- can be quantized by loop quantization
  techniques without requiring any regularization of their
  Hamiltonians or constraints. If such quantizations could be
  performed in general, they would be strongly preferred. Generic
  models, however, suggest that regularization, and thus a role of
  quantum geometry effects for the quantum space-time dynamics, cannot
  be avoided completely.}  motivated from quantum geometry via
background independence; it is not in itself a quantum effect even if,
as sometimes done in improvised versions, $\delta$ is related to
$\hbar$ or $\ell_{\rm P}$ by further ad-hoc arguments.  Indeed, in a
systematic derivation of effective equations describing loop quantum
cosmology, (\ref{RegHam}) is recognized simply as the pure
tree-level contribution where all quantum corrections vanish
\cite{BouncePert,Consistent}; see also
\cite{RegularizationFRW,RegularizationLQC} for discussions of
the regularization. Although care must always be exercised, using just
the regularization allows one to explore the potential consequences of
quantum gravity. The regularization is, of course, easy to implement
in exactly homogeneous models, and it is far from being unique. The
real issues to be faced arise when one tries to extend the
regularization to inhomogeneous situations, at least of perturbative
nature, in which extremely tight constraints due to covariance arise.
Consistent implementations may strongly reduce the ambiguities --- and
possibly eliminate effects seen in simple homogeneous models. Such
issues related to inhomogeneity will be discussed in more detail in
Sec.~\ref{Cosmo}.

An immediate implication of using holonomies is that the constraint
equation is not differential, but a difference equation for a wave
function of the universe \cite{cosmoIV,IsoCosmo}.  Writing
$\hat{C}|\psi\rangle=0$ for a state $|\psi\rangle=
\sum_{\mu}\psi_{\mu}(\phi)|\mu\rangle$ expanded in triad eigenstates
$|\mu\rangle$ (with an extra collective label $\phi$ for matter
fields) requires the coefficients $\psi_{\mu}(\phi)$ to satisfy
\begin{equation} \label{Diff}
    C_+(\mu) \psi_{\mu+\delta}(\phi)- 
C_0(\mu)\psi_{\mu}(\phi)+
  C_-(\mu)\psi_{\mu-\delta}(\phi)
  = \hat{H}_{\phi}(\mu)\psi_{\mu}(\phi)\,.
\end{equation}
All coefficients of this equation can be derived, but are not fixed
uniquely owing to the non-uniqueness of the Hamiltonian constraint
operator.  Nonetheless, several qualitative properties, insensitive to
ambiguities, have been found.  The left-hand side quantizes the
gravitational contribution to the constraint and shows the
discreteness, while the right-hand side shows what role is played by
the matter Hamiltonian $\hat{H}_{\phi}$. If we view the size variable
$\mu$ as an ``internal time,'' evolution proceeds discretely.

Loop quantum cosmology is non-singular \cite{Sing}: any wave function
evolves uniquely across the classical singularity (situated at
$\mu=0$). Quantum hyperbolicity \cite{BSCG} of this form has been
realized not only in isotropic models, but also in anisotropic ones
\cite{HomCosmo,Spin} and even in some inhomogeneous situations such as
spherical symmetry \cite{SphSymmSing}.  Physically, one may explain
this phenomenon by a limited storage for energy provided in a discrete
space-time: Quantum waves must now be supported on a discrete lattice,
providing a natural cut-off for wave-lengths.  Dynamically, a
repulsive force arises once energy densities become too large,
counter-acting the classical attraction and preventing singularities.

In simple models in which also the matter content is severely
restricted by being close to a free, massless scalar --- resulting in
an exactly solvable, harmonic model as shown below ---, numerical
\cite{APS} and exact solutions \cite{BouncePert} indeed show
that the expectation value of the scale factor bounces, reaching a
non-zero minimum value.  This geometrical picture is, however, not
available in strong quantum regimes in which several of the quantum
variables matter: a state changes considerably as the big bang is
approached or traversed, a dynamical behavior which can no longer be
formulated just in terms of the classical variables of geometry
(solely the scale factor in isotropic cosmology). To handle such
situations, effective equations are useful.

\section{Effective equations}

To illustrate the derivation of effective equations in canonical
quantum systems we start with a simple example from quantum mechanics:
the harmonic oscillator. Its dynamics is defined by the closed, linear
algebra
\[
 [\hat{q},\hat{p}]=i\hbar\quad,\quad 
{}[\hat{q},\hat{H}]=i\hbar\frac{\hat{p}}{m}
{}\quad,\quad{}[\hat{p},\hat{H}]=-i\hbar m\omega^2 \hat{q}
\]
of basic operators and the Hamiltonian $\hat{H}$. Any quantum system
with such a closed and linear algebra has dynamical solutions whose
wave functions may spread, but do so without disturbing the mean
position. Indeed, a closed set of equations results for expectation
values of $\hat{q}$ and $\hat{p}$ via ${\rm d}
\langle\hat{O}\rangle/{\rm d} t=
\langle[\hat{O},\hat{H}]\rangle/i\hbar$. To solve these equations, we
need not know how fluctuations, correlations or higher moments of the
state behave; there is no quantum back-reaction. Similarly, there is a
closed set of equations just for the fluctuations and correlations,
without coupling to moments of higher than second order.

A similarly solvable system exists in loop quantum cosmology
\cite{BouncePert,BounceCohStates}, with the conditions of an
isotropic, spatially flat space and matter given by a free, massless
scalar $\phi$. Using the loop-quantized Hamiltonian (in a particular
factor ordering and ignoring inverse volume corrections at this stage)
again produces a closed algebra of basic variables, provided we choose
them as the volume $V$ (or, more generally, $|p|^{1-x}$ to absorb any
$p$-dependence of $\delta$ provided it is a power-law
$\delta(p)=\delta_0 |p|^x$) and the holonomy-related quantity
$J=V\exp(i\delta_0 P)$ with the Hubble parameter $P$ (or $|p|^xc$)
conjugate to $V$. This time, using the Hamiltonian $p_{\phi}\propto h:=
{\rm Im}J$ with respect to evolution in $\phi$ as internal time, as it
follows from the regularized Hamiltonian (\ref{RegHam}) with
$H_{\rm matter}= p_{\phi}^2/2a^3$, the algebra is sl$(2,{\mathbb R})$:
\[
 [\hat{V},\hat{J}]=-\delta_0\hbar\hat{J}\quad,\quad{}
 [\hat{V},\hat{h}]={\textstyle\frac{1}{2}}i\delta_0\hbar
 (\hat{J}+\hat{J}^{\dagger})\quad,\quad{}
 [\hat{J},\hat{h}]=i\delta_0\hbar\hat{V}\,.
\]
(The Hamiltonian constraint for a free, massless scalar field can be
written as $p_{\phi}^2-VC_{\rm grav}=0$, and easily be
deparameterized. One is taking a square root in the process to solve
for $p_{\phi}$, but this does not spoil the linearity of the dynamics
of states just required to be semiclasical once
\cite{BounceCohStates}. Alternatively, direct treatments of effective
constraints, avoiding deparameterization, are available
\cite{EffCons,EffConsRel,EffConsComp}.) 

\begin{figure}[t]
\centering
\includegraphics[keepaspectratio=true,width=9cm]{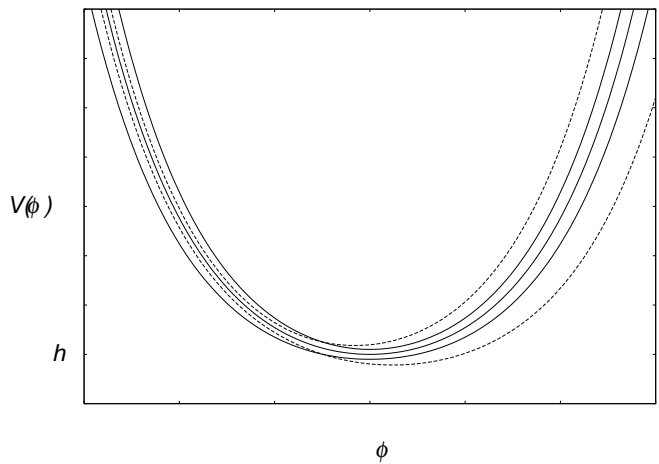}
\caption{Bouncing solutions of the exactly solvable model of loop
  quantum cosmology, showing the expectation value of the physical
  observable $\langle\hat{V}\rangle(\phi)$, as well as the spread
  $\Delta V(\phi)$of a physical wave function. The label ``$h$''
  indicates the size of the (conserved) $\phi$-Hamiltonian, which
  determines the volume at the bounce.}
\label{fig:EffBounce}
\end{figure}

Equations of motion
\[
\frac{{\rm d} \langle\hat{O}\rangle}{{\rm d}\phi}=
\frac{\langle[\hat{O},\hat{h}]\rangle}{i\hbar}
\]
generated with respect to $\phi$ now provide the behavior of physical
observables: There is no absolute time in this fully constrained
system; instead, change is measured by relational observables, such as
$\langle\hat{V}\rangle(\phi)$ between the degrees of freedom. (For a
complete reduction to physical quantities, reality conditions must be
imposed to ensure the correct adjointness properties for a
quantization of the real $P$ appearing in the complex $J$. Appropriate
conditions turn out to be easily formulated, relating expectation
values to fluctuations and correlations \cite{BounceCohStates}.)
Also here, there is no quantum back-reaction in the solvable model.
Fluctuations do not back-react on the expectation values, which
results in simple, cosh-like solutions for the volume; see
Fig.~\ref{fig:EffBounce}.  Clearly, the volume never becomes zero,
and the singularity, of these specific models, is replaced by a
bounce. While the expectation value follows its trajectory
undisturbed, states in general do spread. In particular, squeezed
states (with non-vanishing correlations) describe oscillating
fluctuations between different universe phases, expansion and
collapse. As it turns out, fluctuations can change by an order of
magnitude even in dynamical coherent states --- the most strongly
controlled type of states ---, and this change is very sensitive to
initial values. This cosmic forgetfulness makes it difficult to
estimate specific quantities in the pre-bounce phase for realistic
models \cite{BeforeBB,Harmonic}.

Generic models are more complicated since they are subject to quantum
back-reaction. To illustrate this, we consider a model with a
cosmological constant, but ignore the quantum geometry corrections of
loop quantum cosmology for the sake of simplicity. The Hamiltonian for
$\phi$-evolution, as treated for a negative cosmological constant in
\cite{Recollapse}, then is $p_{\phi} \propto V\sqrt{P^2-\Lambda} =:
h(V,P)$. (The same system was analyzed numerically in
\cite{NegCosNum}.)  Now, expectation values couple to fluctuations and
other moments:
\begin{eqnarray*}
 \frac{{\rm d} \langle\hat{V}\rangle}{{\rm d}\phi} &=& \frac{3}{2}\frac{\langle\hat{V}\rangle\langle\hat{P}\rangle}{\sqrt{\langle\hat{P}\rangle^2-\Lambda}}
+\frac{9}{4}\Lambda \frac{\langle\hat{V}\rangle\langle\hat{P}\rangle}{(\langle\hat{P}\rangle^2-\Lambda)^{5/2}} (\Delta P)^2
-\frac{3}{2}\Lambda \frac{C_{VP}}{(\langle\hat{P}\rangle^2-\Lambda)^{3/2}}+\cdots \\ 
\frac{{\rm d} \langle\hat{P}\rangle}{{\rm d}\phi} &=& -\frac{3}{2}\sqrt{\langle\hat{P}\rangle^2-\Lambda}+ \frac{3}{4}\Lambda
\frac{(\Delta P)^2}{(\langle\hat{P}\rangle^2-\Lambda)^{3/2}}+\cdots
\end{eqnarray*}
with the $\hat{P}$-fluctuation $\Delta P$ and the covariance $C_{VP}=
\frac{1}{2}\langle\hat{V}\hat{P}+\hat{P}\hat{V}\rangle-
\langle\hat{V}\rangle\langle\hat{P}\rangle$.  Fluctuations are
dynamical, too:
\begin{eqnarray*}
 \frac{{\rm d} (\Delta P)^2}{{\rm d}\phi} &=& -3\frac{\langle\hat{P}\rangle}{\sqrt{\langle\hat{P}\rangle^2-\Lambda}} (\Delta P)^2+\cdots\\
 \frac{{\rm d} C_{VP}}{{\rm d}\phi} &=& -\frac{3}{2}\Lambda
\frac{\langle\hat{V}\rangle}{(\langle\hat{P}\rangle^2-\Lambda)^{3/2}}(\Delta P)^2+\cdots \\
 \frac{{\rm d} (\Delta V)^2}{{\rm d}\phi} &=& -3 \Lambda
\frac{\langle\hat{V}\rangle}{(\langle\hat{P}\rangle^2-\Lambda)^{3/2}} C_{VP} 
 +3\frac{\langle\hat{P}\rangle}{\sqrt{\langle\hat{P}\rangle^2-\Lambda}} (\Delta V)^2+\cdots\,.
\end{eqnarray*}
(In all these equations, dots indicate that higher order moments have
been ignored here.)

Analyzing coupled equations like this is the canonical procedure for
effective equations. (For anharmonic oscillators in quantum mechanics,
employing a semiclassical as well as an adiabatic approximation, the
usual low-energy effective action is reproduced
\cite{EffAc,Karpacz}.)  Often, higher moments can be ignored in
certain regimes, starting with a semiclassical state whose moments of
order $n$ are suppressed by a factor of $\hbar^{n/2}$. But long
evolution, as is prevalent in cosmology, can drastically change a
state even if it starts out semiclassically to a high degree. Moments
may then grow, and higher ones will become important. Severe quantum
back-reaction effects can be expected, especially in the infamously
strong quantum regime around the big bang. But also other regimes
exist where large moments are perhaps more surprising. One example is
that of the large-volume regime of models with a positive cosmological
constant.  As can be seen from the preceding equations, several of the
coefficients then have denominators that can come close to zero when
the curvature scale squared approaches the cosmological constant. For
a small, perhaps realistic, cosmological constant, this regime is
approached at large volume, where moments can grow large despite the
classical appearance of the phase.

The back-reaction equations of loop quantum cosmology are much more
lengthy \cite{BouncePot}, but can be summarized in a quantum Friedmann
equation including the effects from holonomy corrections and quantum
back-reaction \cite{QuantumBounce,BounceSqueezed}.  With a scalar mass
or a self-interacting potential, the equation
\begin{equation}
 \left(\frac{\dot{a}}{a}\right)^2 = \frac{8\pi G}{3}\left(\rho
 \left(1-\frac{\rho_Q}{\rho_{\rm crit}}\right)
 \pm\frac{1}{2}\sqrt{1-\frac{\rho_Q}{\rho_{\rm crit}}} 
\eta (\rho-P)+ \frac{(\rho-P)^2}{(\rho+P)^2}\eta^2
\right)
\end{equation}
describes the evolution of the scale factor's expectation
value. Compared to the classical equation, pressure $P$ enters, as
well as $\eta$ which parameterizes quantum correlations. Moreover,
\[
 \rho_Q:=\rho+\epsilon_0 \rho_{\rm crit}+ (\rho-P) \sum_{k=0}^{\infty}
\epsilon_{k+1}
 (\rho-P)^k/(\rho+P)^k
\]
is an expression for a quantum corrected energy density with
fluctuation parameters $\epsilon_k$, and $\rho_{\rm crit}=3/8\pi
G(a\delta)^2$ is a critical density with scale $a\delta$ as determined
by the $\delta$ used in the regularization (\ref{RegHam}) by
holonomies. The critical density is constant only if $\delta\propto
a^{-1}$ (a special case introduced in \cite{APSII}).  The behavior
following from this equation is simple if $\rho=p$ (the free, massless
scalar case) or if $\rho+p$ is large (large $p_{\phi}$, or kinetic
domination). Then, solutions with $\dot{a}=0$ exist near
$\rho=\rho_{\rm crit}$, producing a bounce. 

For regimes not of kinetic domination, the behavior of many moments,
contained in $\eta$ and $\rho_Q$, must be known for a precise picture,
requiring a long analysis still to be completed. Only such an analysis
can show what effective geometrical picture corresponds to the general
singularity avoidence by the difference equation of loop quantum
cosmology. In particular, it has {\em not} been shown that loop
quantum cosmology generically replaces the big bang singularity by a
bounce.

Currently, all existing indications for bounces --- numerical as in
\cite{QuantumBigBang,InhomThroughBounce} or analytic as in
\cite{BounceSqueezed} --- exist only for kinetic-dominated regimes of
a scalar matter source. The situation is slightly more general for
demonstrating an upper bound of energy density, but such a statement
is weaker than showing the existence of bouncing solutions. (Bounces
can easily be produced quite generally even in potential-dominated
regimes using the tree-level approximation of loop quantum cosmology.
However, the tree-level approximation itself does not appear reliable
in potential-dominated regimes.)

\section{Lattice refinement}
\label{sec:LattRef}

Loop quantum cosmology aims to model the dynamical behavior of loop
quantum gravity in a tractable manner. Since no procedure is known for
a complete reduction of the Hamiltonian constraint to isotropy or
homogeneity, several choices are to be made in specifying the
Hamiltonian constraint of loop quantum cosmology, giving rise to the
difference equation (\ref{Diff}). One such freedom concerns the
parameter $\delta$ in (\ref{RegHam}), which may be a phase-space
function as alluded to above.

This function carries important information about the reduction
\cite{InhomLattice,CosConst}. It appears as a coefficient of the
connection component in holonomies as used for the dynamics. If we
look at the schematic full constraint of (\ref{HamOp}), it is clear
that holonomies in that operator refer to edges in one of the graph
states, as it evolves according to the dynamics of loop quantum
gravity. Reducing such a holonomy to isotropic variables leads to an
expression of the form $\exp(i\delta c)$, exactly as used in the
reduced constraint. In the reduced model, $\delta$ appears as a
parameter which can only be chosen by hand to have a specific value;
no argument has been found to fix it. In the full context, on the
other hand, $\delta$ is clearly related to the coordinate length of
the edge used, and thus refers to the underlying inhomogeneous state.
(Although $\delta$ is coordinate dependent, the combination $\delta
c=\gamma\delta\dot{a}$ appearing in holonomies is not.) In the
reduction to homogeneity, that information in the state is lost; one
can only bring it back by making certain phenomenological choices for
$\delta$.

The underlying inhomogeneous state is dynamical: new edges may be
created or old ones removed. Edge lengths change, and so does
$\delta$. One way to model this in an isotropic description is to
allow $\delta$ to depend on the scale factor $a$, implying that the
underlying inhomogeneous state changes as the universe expands or
contracts. By analyzing the resulting models, phenomenological
restrictions for the behavior of $\delta$ can be found
\cite{RefinementMatter,RefinementInflation,TensorSlowRoll}.
What is so far indicated is that a power-law form of
$\delta=\delta_0|p|^x$ works best for $x$ near $-1/2$.

Lattice refinement has been formulated in a parameterized way for
anisotropic models, also at the level of underlying difference
equations. Compared to isotropic models, the difference equations then
generically become non-equidistant, complicating an analysis. A
complete formulation of the dynamics (avoiding ad-hoc assumptions),
especially for the Schwarzschild interior but also providing
Misner-type variables for Bianchi models, has been provided in
\cite{SchwarzN}. Numerical tools to evaluate non-equidistant
difference equations have been introduced in
\cite{RefinedNumeric,RefinementNumeric}.

\section{Cosmology}
\label{Cosmo}

With matter interactions and inhomogeneities, a complicated form of
back-reaction results that can be handled only by a systematic
perturbation theory around the solvable model. The solvable model of
loop quantum cosmology then plays the same role as free quantum field
theories do for the Feynman expansion.  An analysis of this form can
show possible indirect effects of the atomic space-time where
individual corrections which are small even at high energies might add
up coherently. If this magnification effect is strong enough, one
might come close to observability. Two prime examples exist:
cosmology, which is a high energy density regime with long evolution
times for corrections to add up; and high energy particles from
distant sources.

But before one analyzes complete equations --- those containing all
possible quantum corrections for possible physical consequences ---
there is an interesting geometrical set of problems related to general
covariance.  Compared to homogeneous models, where modifications such
as those in (\ref{RegHam}) can consistently be implemented at will,
general covariance in inhomogeneous situations is a strong consistency
requirement. For instance, the contracted Bianchi identity
$\nabla_{\mu} G^{\mu}_{\nu}=0$ implies $\partial_0G^0_{\mu}=
-\partial_aG^a_{\mu}-\Gamma^{\nu}_{\nu\kappa} G^\kappa_{\mu}+
\Gamma^{\kappa}_{\mu\nu}G^{\nu}_{\kappa}$\,.  The right-hand side is
at most of second order in time derivatives, and so $G^0_{\mu}$ must
be of first order. The corresponding components of Einstein's equation
provide constraints for initial values rather than evolution
equations:
\begin{equation}
 \int{\rm d}^3x N(x) \sqrt{\det q} (G^0_0-8\pi GT^0_0)=0
\end{equation}
(with the spatial metric tensor $q_{ab}$ used in the integration
measure) and
\begin{equation}
  \int{\rm d}^3x N^a(x)\sqrt{\det q}(G^0_a-8\pi G T^0_a)=0
\end{equation}
where $q_{ab}$ is again the spatial metric. The constraints must be
preserved under the second order equations that follow from the
spatial components $G^a_b$.

This kind of conservation law leads to symmetries: We have a scalar
constraint $C[N]$, the Hamiltonian constraint, and a vectorial one,
the diffeomorphism constraint $D[N^a]$, satisfying a closed algebra
\begin{eqnarray*}
 \{D[N^a],D[M^a]\}&=& D[{\cal L}_{M^a}N^a]\\
 \{C[N],D[M^a]\} &=& C[{\cal L}_{M^a}N]\\
 \{C[N],C[M]\} &=& D[q^{ab}(N\partial_bM-M\partial_bN)]
\end{eqnarray*}
as the generators of gauge transformations. Importantly, this algebra
is first class: Poisson brackets of the constraints vanish when the
constraints are imposed. On the solution space, constraints are
invariant under the flow they generate, and thus provide
gauge-invariant equations.  In the case of gravity, the combination
$C[N]+D[N^a]$ generates infinitesimal space-time diffeomorphism along
$\xi^{\mu}=(N,N^a)$, as can be checked by a direct calculation and
comparison with Lie derivatives. General covariance can be expressed
fully in terms of this algebra, as emphasized by Dirac
\cite{DiracHamGR}:
\begin{quote}
  ``It would be permissible to look upon the Hamiltonian form as the
  fundamental one, and there would then be no fundamental
  four-dimensional symmetry in the theory.

  The usual requirement of four-dimensional symmetry in physical laws
  would then get replaced by the requirement that the functions have
  weakly vanishing [Poisson brackets].''
\end{quote}

Such a viewpoint is convenient especially in canonical quantum
gravity. It is not guaranteed that quantization preserves the usual
space-time or differential-geometric notions, and that it leads to the
same relationship between symmetries and Lie derivatives. In contrast
to differential geometry, an algebra of constraints, which will become
a commutator algebra of the corresponding operators, can directly be
carried over to the quantum theory. In this way, the realization of
symmetries, and correspondingly of space-time structures, can be
tested at the quantum level. Quantum corrections usually change the
constraints as gauge generators and may thus lead to changes in the
space-time structures.  Also the algebra of constraints may be
corrected, but for a consistent formulation, corrections must respect
the first-class nature of the algebra. If the first-class nature is
respected, symmetries may be deformed but are not lost; the quantum
system is then called anomaly-free.

``Effective'' constraints, including some of the corrections from
quantum geometry (in this case inverse volume corrections), can be
made anomaly-free \cite{ConstraintAlgebra}:
\begin{eqnarray*}
 \{D[N^a],D[M^a]\}&=& D[{\cal L}_{M^a}N^a]\\
 \{C_{(\alpha)}[N],D[M^a]\} &=& C_{(\alpha)}[{\cal L}_{M^a}N]\\
 \{C_{(\alpha)}[N],C_{(\alpha)}[M]\} &=& 
D[\alpha^2 q^{ab}(N\partial_bM-M\partial_bN)]
\end{eqnarray*}
where $\alpha$ is the correction function from inverse volume
operators. This algebra is indeed first class, but deformed. To
interpret the corrections, we first note that in an effective action
they cannot be purely of higher curvature type, for such corrections
would still produce the classical algebra \cite{HigherCurvHam}. We
are thus dealing with a more general type of effective action (such as
one on a non-commutative space-time). Similar deformations have been
constructed for holonomy corrections, although not for the complete
case of cosmological perturbations. The first example was found for
spherically symmetric models \cite{JR} (see also
\cite{LTB,LTBII}), with a similar form produced for
$2+1$-dimensional gravity \cite{ThreeDeform}. (Although the
deformations in those two cases are quite similar, there is a
difference in that the $2+1$-example requires a non-vanishing
cosmological constant for the deformation to appear. This circumstance
may just be a consequence of the special form of $2+1$-dimensional
gravity in the formulation used, where the theory without a
cosmological constant has vanishing on-shell curvature and is
topological.) Constructing consistent deformations corresponding to
quantum geometry corrections is the non-trivial part of an analysis
making simple modifications such as (\ref{RegHam}) relevant.

Practically, one consequence of the deformation is that the potential
size of quantum corrections is larger than often expected, that is
larger than $\ell_{\rm P}{\cal H}$ as higher curvature terms would
produce it in cosmological situations.  The main physical mechanism is
non-conservation of power on large scales, modifying an approximate
conservation which classically follows very generally from the
conservation of stress-energy, or the Bianchi identity; see e.g.\
\cite{SeparateUniverse,SeparateUniverseII,SeparateUniverseIII}.
The Bianchi identity, however, takes a different form for the
corrected constraint algebra, and so quantum corrections affect
large-scale modes, removing the constant one
\cite{InhomEvolve,ScalarGaugeInv}.  Local corrections for the
slope of increase or decrease of super-Hubble power, for scalar and
tensor modes, are small, but realized during long evolution times.
They may add up to sizeable effects.

Explicitly, the resulting cosmological perturbation equations (for all
scales on a background Friedmann--Robertson--Walker space-time with
conformal Hubble rate ${\cal H}$ and the background scalar
$\bar{\varphi}$) are \cite{ScalarGaugeInv}
\begin{equation}
\partial_c\left(\dot\Psi+{\cal H}(1+f)\Phi\right)=4\pi
G\frac{\bar{\alpha}}{\bar{\nu}}\dot{\bar{\varphi}} \partial_c\delta\varphi
\end{equation}
from the diffeomorphism constraint,
\begin{equation}
\Delta(\bar{\alpha}^2
\Psi)-3{\cal H}(1+f)
\left(\dot\Psi+{\cal H}\Phi(1+f)\right)
=4\pi
G\frac{\bar{\alpha}}{\bar{\nu}}(1+f_3)
\left(\dot{\bar{\varphi}}\delta\dot\varphi-\dot{\bar{\varphi}}^2(1+f_1)\Phi
+\bar{\nu} a^2 V_{,\varphi}(\bar{\varphi})
\delta\varphi\right)
\end{equation}
from the Hamiltonian constraint, and
\begin{eqnarray}
&&\ddot\Psi+{\cal H}\left(2\dot\Psi\left(1-\frac{a}{2\bar{\alpha}}
\frac{{\rm d}\bar{\alpha}}{{\rm d}a}\right)+\dot\Phi(1+f)\right)
+\left(2\dot{\cal H}+{\cal H}^2\left(1+
\frac{a}{2}\frac{{\rm d}f}{{\rm d}a} -
\frac{a}{2\bar{\alpha}}\frac{{\rm d}\bar{\alpha}}{{\rm d}a}\right)\right)
\Phi(1+f)\nonumber\\
&=&4\pi G\frac{\bar{\alpha}}{\bar{\nu}}
\left(\dot{\bar{\varphi}}\delta\dot\varphi-a^2\bar{\nu} V_{,\varphi}(\bar{\varphi})\delta\varphi\right)
\end{eqnarray}
as the evolution equation. These equations are accompanied by
$\Phi=(1+h)\Psi$, which follows from off-diagonal components of the
corrected Einstein equation, and a corrected Klein--Gordon equation
for $\delta\varphi$.  In addition to the gauge-invariant metric
perturbations $\Phi$ and $\Psi$ and the scalar perturbation
$\delta\varphi$ as well as the primary correction function
$\bar{\alpha}$ from inverse volume (and $\bar{\nu}$ for the matter
term), several other corrections, $f$, $f_1$, $f_3$, $h$ arise, but
are related to the primary correction.

The consistency issue now becomes a very practical problem: There are
five equations for three free functions, $\Phi$, $\Psi$ and
$\delta\varphi$.  Classically the system is consistent and not
overdetermined thanks to general covariance. But will this closure of
the equations be preserved in the presence of quantum corrections from
discrete geometry? As indicated by the possibility of a first-class
(but deformed) algebra of constraints, consistency can be realized.
There are no anomalies (checked to linear order in perturbations in
\cite{ConstraintAlgebra}) if the quantum correction functions satisfy
equations such as
\begin{eqnarray*}
 -h -f+\frac{a}{\bar{\alpha}}\frac{\partial
\bar{\alpha}} {\partial a} &=&0\quad,\quad
3f-2a\frac{\partial f}{\partial a}
-\frac{a}{\bar{\alpha}}\frac{\partial \bar{\alpha}}
{\partial a} =0 \\
\frac{1}{6}\frac{\partial\bar{\alpha}}{\partial a}
\frac{\delta E^c_j}{a^3}
+ \frac{\partial\alpha^{(2)}}{\partial(\delta E^a_i)}
(\delta^a_j \delta^c_i - \delta^c_j \delta^a_i) &=& 0\,.
\end{eqnarray*}
The last of these equation relates higher perturbative orders of
$\alpha$ to the background value $\bar{\alpha}$ achieved for isotropic
geometries.  If these equations are satisfied, which is possible even
in non-classical cases of $\alpha\not=1$, the whole set of
cosmological perturbation equations is consistent. Moreover, all
coefficients for quantum corrections are fixed in terms of
$\bar{\alpha}$, and this function can be derived in isotropic models
(up to ambiguities).  Inverse volume corrections, resulting from
discrete features of spatial quantum geometry, provide a consistent
deformation: The underlying discreteness (for this type of
corrections) does not destroy general covariance.

General covariance is a statement about the quantum constraint
algebra, and thus can be ensured only by considering perturbation
theory without fixing the space-time gauge. After consistent equations
have been derived, one may certainly pick a gauge (such as the
longitudinal one) if that simplifies calculations. But fixing the
gauge before deriving perturbation equations and observables does not
verify covariance and can easily lead to spurious effects. Explicit
examples can be seen by comparing \cite{HamPerturb} with
\cite{ScalarGaugeInv}, the first one with gauge-fixing, the second one
without. As turns out, the relationship between the metric
perturbations $\Phi$ and $\Psi$ is affected by the treatment, as is
the precise form of non-conserved power on large scales. Only without
gauge-fixing is it possible to ensure consistency; only those results
are reliable. When computing the effects of an underlying discrete
space-time structure on inflationary structure formation, or on the
propagation of modes through a potential bounce, that kind of
consistency is especially important. For evolution through a bounce,
no consistent form for scalar modes has been found yet; existing
treatments all use gauge-fixing
\cite{HolonomyInfl,BounceCMB,CosPertHolLong}.

\begin{figure}[t]
\centering
\includegraphics[keepaspectratio=true,width=9cm]{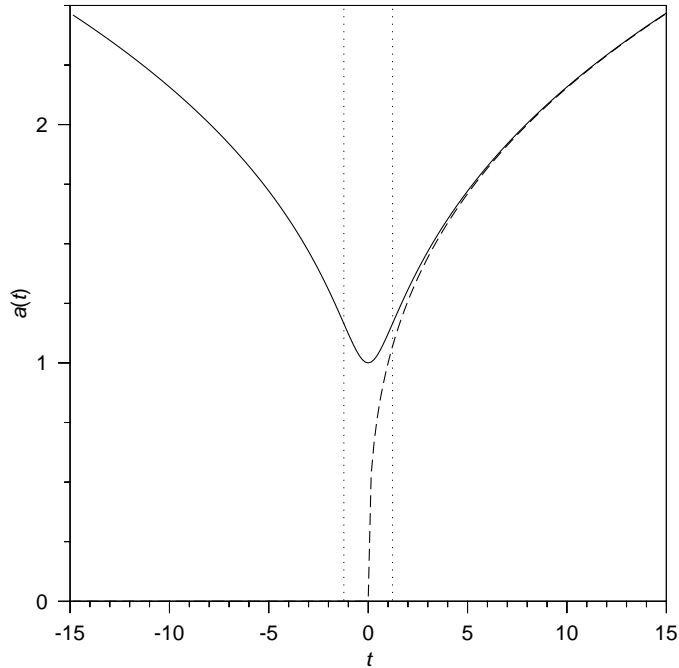}
\caption{A bouncing scale factor, with super-inflation between the
  dotted lines (and the singular classical solution as the dashed
  line). In this phase, the scale factor varies only by a few percent,
  but $a{\cal H}$ changes more. The effective solution corresponds to
  the solvable model as plotted in Fig.~\ref{fig:EffBounce},
  transformed to proper time.  Similar solutions have been found in
  \cite{BounceSols} directly in proper time.}
\label{fig:Bounceac}
\end{figure}

One of the implications for cosmological scenarios is that quantum
geometry corrections (inverse volume or holonomy) often imply
super-inflation at high densities \cite{Inflation}. There may not
be many e-foldings in terms of $\log(a_{\rm f}/a_{\rm i})$, referring
only to the final and initial values of the scale factor (see
Fig.~\ref{fig:Bounceac}), but $\log\left((a{\cal H})_{\rm
    f}/(a{\cal H})_{\rm i}\right)$ can be large due to the growth of
${\cal H}$ during super-inflation \cite{SuperInflLQC}. Viable
scenarios thus exist, but the degree of fine-tuning has not yet been
fully estimated. Regarding details, several analyses have been
performed depending on the scalar potential and quantization
ambiguities.  Unfortunately, perturbation results are currently on a
rather weak basis in strong quantum (or high density) regimes since
anomaly-free effective equations are difficult to control. For
instance, no consistent evolution of scalar modes through the bounce
yet exists.  Nevertheless, in weaker regimes some parameters, such as
the power-law of a scalar potential, can already be constrained (in
\cite{SuperInflPowerSpec}, using WMAP5 data). At the current stage,
primordial gravitational waves \cite{tensor} are under better
analytical control since they are not subject to gauge transformation
or overdeterminedness. Some implications for the tensor power spectrum
have been evaluated in
\cite{TensorBackground,TensorRelic,TensorHalf,TensorHalfII,BounceTensor,TensorSlowRollInv,TensorSlowRoll}.

\section{Outlook}

Consistent deformations exist in model systems of canonical quantum
gravity: discreteness can be realized without spoiling covariance.
These results of anomaly freedom show that discrete structures are
able to preserve covariance; when realized, they make simple
modifications as they are possible in homogeneous models highly
non-trivial. So far, this demonstration has been achieved only in
relatively tame regimes, but not yet close to the classical big bang
singularity, or even through bounce phases.

Examples have been constructed for cosmological perturbations and for
black holes. This is mainly based on canonical effective equations,
whose tools, analytical as well as numerical ones, are currently being
developed. In some cases, these equations already provide a link to
cosmological, astrophysical and particle observations. One general
result seems to be that the quantum space-time structure is certainly
important in high-energy regimes, but, thanks to magnification
effects, not necessarily just at the Planck scale. This allows one to
set bounds on parameters of quantum space-time, an extensive
investigation that is still ongoing.

\section*{Acknowledgements}

The author is grateful to Tomohiro Harada for an invitation to The
Nineteenth Workshop on General Relativity and Gravitation in Japan
(JGRG19) at Rikkyo University.
Work reported here was supported by NSF grant PHY0748336 and a grant
from the Foundational Questions Institute (FQXi).

\end{document}